\documentclass{iaus}
\usepackage{graphicx}
\title[The First Stars] %% give here a short title %%
{The First Stars}

\author[Johnson, Greif \& Bromm]   %% give here a short author list %%
{Jarrett L. Johnson$^1$,Thomas H. Greif $^2$ \break \and Volker Bromm$^1$}

\affiliation{$^1$Department of Astronomy, University of Texas, Austin, TX 78712 \\[\affilskip]
$^2$Institut f\"{u}r Theoretische Astrophysik, \break Albert-Ueberle Strasse 2, 69120 Heidelberg, Germany}

\pubyear{2007}
\volume{250}  %% insert here the IAU Symposium No.
\pagerange{1--7}
\date{?? and in revised form ??}
\jname{??}
\editors{}
\begin{document}

\maketitle

\begin{abstract}
The formation of the first generations of stars at redshifts $z\geq 15-20$ signaled the transition from the simple initial state of the universe to one of ever increasing complexity. We here review recent progress in understanding the assembly process of the first galaxies, starting with cosmological initial conditions and modelling the detailed physics of star formation. In particular, we study the role of HD cooling in ionized primordial gas, the impact of UV radiation produced by the first stars, and the propagation of the supernova blast waves triggered at the end of their brief lives. We conclude by discussing how the chemical abundance patterns observed in extremely low-metallicity stars allow us to probe the properties of the first stars.
%% add here a maximum of 10 keywords, to be taken form the file <Keywords.txt>
\end{abstract}

\firstsection % if your document starts with a section,
              % remove some space above using this command.
\section{Introduction}
 One of the key goals in modern cosmology is to study the formation of the first generations of stars and to understand the assembly process of the first galaxies. With the formation of the first stars, the so-called Population~III (Pop~III), the universe was rapidly transformed into an increasingly complex, hierarchical system, due to the energy and heavy element input from the first stars and accreting black holes (Barkana \& Loeb 2001; Bromm \& Larson 2004; Ciardi \& Ferrara 2005; Miralda-Escud{\'e} 2003). Currently, we can directly probe the state of the universe roughly a million years after the Big Bang by detecting the anisotropies in the cosmic microwave background (CMB), thus providing us with the initial conditions for subsequent structure formation. Complementary to the CMB observations, we can probe cosmic history all the way from the present-day universe to roughly a billion years after the Big Bang, using the best available ground- and space-based telescopes. In between lies the remaining frontier, and the first stars and galaxies are the sign-posts of this early, formative epoch.

To simulate the build-up of the first stellar systems, we have to address the feedback from the very first stars on the surrounding intergalactic medium (IGM), and the formation of the second generation of stars out of material that was influenced by this feedback. There are a number of reasons why addressing the feedback from the first stars and understanding second-generation star formation is crucial:\\ {\it (i)} The first steps in the hierarchical build-up of structure provide us with a simplified laboratory for studying galaxy formation, which is one of the main outstanding problems in cosmology. 
\\ {\it (ii)} The initial burst of Pop~III star formation may have been rather brief due to the strong negative feedback effects that likely acted to self-limit this formation mode (Greif \& Bromm 2006; Yoshida et al. 2004). Second-generation star formation, therefore, might well have been cosmologically dominant compared to Pop~III stars.\\ {\it (iii)} A subset of second-generation stars, those with masses below $\simeq 1~M_{\odot}$, would have survived to the present day. Surveys of extremely metal-poor Galactic halo stars therefore provide an indirect window into the Pop~III era by scrutinizing their chemical abundance patterns, which reflect the enrichment from a single, or at most a small multiple of, Pop~III SNe (Beers \& Christlieb 2005; Frebel et al. 2007). Stellar archaeology thus provides unique empirical constraints for numerical simulations, from which one can derive theoretical abundance patterns to be compared with the data.

Existing and planned observatories, such as HST, Keck, VLT, and the {\it James Webb Space Telescope (JWST)}, planned for launch around 2013, yield data on stars and quasars less than a billion years after the Big Bang. The ongoing {\it Swift} gamma-ray burst (GRB) mission provides us with a possible window into massive star formation at the highest redshifts (Bromm \& Loeb 2002, 2006; Lamb \& Reichart 2000). Measurements of the near-IR cosmic background radiation, both in terms of the spectral energy distribution and the angular fluctuations provide additional constraints on the overall energy production due to the first stars (Dwek et al. 2005; Fernandez \& Komatsu 2006; Kashlinsky et al. 2005; Magliocchetti et al. 2003; Santos et al. 2002). Understanding the formation of the first galaxies is thus of great interest to observational studies conducted both at high redshifts and in our local Galactic neighborhood.

\begin{figure}
\centering
\includegraphics[width=.45\textheight]{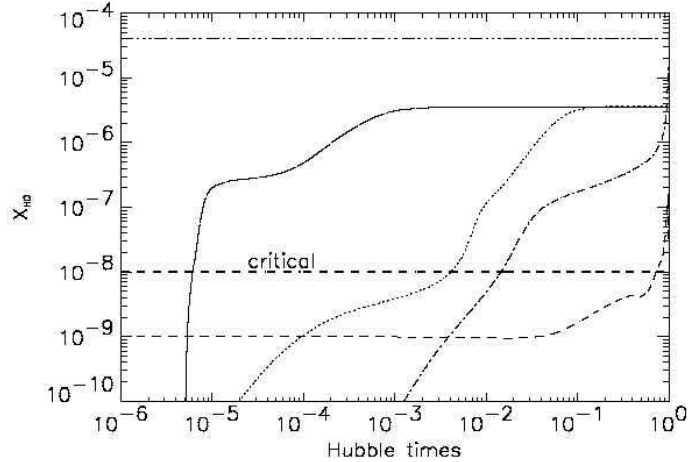}
\caption{Evolution of the HD abundance, $X_{\rm{HD}}$, in primordial gas which cools in four distinct situations. The solid line corresponds to gas with an initial density of $100~\rm{cm}^{-3}$, which is compressed and heated by a SN shock with velocity  $v_{\rm{sh}}=100~\rm{km}~\rm{s}^{-1}$ at $z=20$. The dotted line corresponds to gas at an initial density of $0.1~\rm{cm}^{-3}$ shocked during the formation of a $3\sigma$ halo at $z=15$. The dashed line corresponds to unshocked, un-ionized primordial gas with an initial density of $0.3~\rm{cm}^{-3}$ collapsing inside a minihalo at $z=20$. Finally, the dash-dotted line shows the HD fraction in primordial gas collapsing from an initial density of $0.3~\rm{cm}^{-3}$ inside a relic H~{\sc ii} region at $z=20$. The horizontal line at the top denotes the cosmic abundance of deuterium. Primordial gas with an HD abundance above the critical value, $X_{\rm{HD,crit}}$, denoted by the bold dashed line, can cool to the CMB temperature within a Hubble time.}
\end{figure}

\section{Population III Star Formation}

The first stars in the universe likely formed roughly 150 Myr after the Big Bang, when the primordial gas was first able to cool and collapse into dark matter minihalos with masses of the order of 10$^6$ $M_{\odot}$ (Bromm et al. 1999, 2002; Abel et al. 2002).  These stars are believed to have been very massive, with masses of the order of 100 $M_{\odot}$, owing to the limited cooling properties of primordial gas, which could only cool in minihalos through the radiation from H$_2$ molecules.  While the initial conditions for the formation of these stars are, in principle, known from precision measurements of cosmological parameters (e.g. Spergel et al. 2007), Pop~III star formation may have occurred in different environments which may allow for different modes of star formation.  Indeed, it has become evident that Pop~III star formation might actually consist of two distinct modes: one where the primordial gas collapses into a DM minihalo (see below), and one where the metal-free gas becomes significantly ionized prior to the onset of gravitational runaway collapse (Johnson \& Bromm 2006). We had termed this latter mode of primordial star formation `Pop~II.5' (Greif \& Bromm 2006; Johnson \& Bromm 2006; Mackey et al. 2003). To more clearly indicate that both modes pertain to {\it metal-free} star formation, we here follow the new classification scheme suggested by Chris McKee (see McKee \& Tan 2007; Johnson et al. 2008). Within this scheme, the minihalo Pop~III mode is now termed Pop~III.1, whereas the second mode (formerly `Pop~II.5') is now called Pop~III.2. The hope is that McKee's terminology will gain wide acceptance.

\begin{figure}
\centering
\includegraphics[width=.35\textheight]{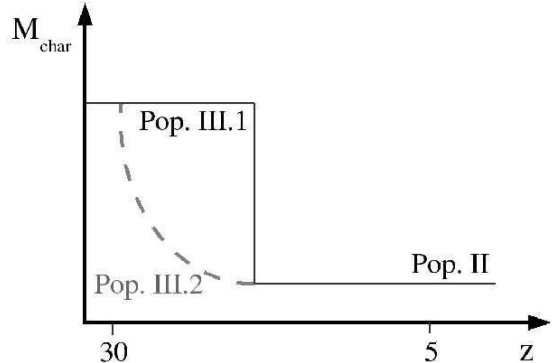}
\caption{Characteristic stellar mass as a function of redshift. Pop~III.1 stars, formed from unshocked, un-ionized primordial gas are characterized by masses of the order of $100~M_{\odot}$. Pop~II stars, formed in gas which is enriched with metals, emerged at lower redshifts and have characteristic masses of the order of $1~M_{\odot}$. Pop~III.2 stars, formed from ionized primordial gas, have characteristic masses reflecting the fact that they form from gas that has cooled to the temperature of the CMB. Thus, the characteristic mass of Pop~III.2 stars is a function of redshift, but is typically of the order of $10~M_{\odot}$.}
\end{figure}

While the very first Pop~III stars (so-called Pop III.1), with masses of the order of $100~M_{\odot}$, formed within DM minihalos in which primordial gas cools by H$_2$ molecules alone, the HD molecule can play an important role in the cooling of primordial gas in several situations, allowing the temperature to drop well below $200~\rm{K}$ (Abel et al. 2002; Bromm et al. 2002). In turn, this efficient cooling may lead to the formation of primordial stars with masses of the order of $10~M_{\odot}$ (so-called Pop III.2) (Johnson \& Bromm 2006). In general, the formation of HD, and the concomitant cooling that it provides, is found to occur efficiently in primordial gas which is strongly ionized, owing ultimately to the high abundance of electrons which serve as catalyst for molecule formation in the early universe (Shapiro \& Kang 1987).

Efficient cooling by HD can be triggered within the relic H~{\sc ii} regions that surround Pop~III.1 stars at the end of their brief lifetimes, owing to the high electron fraction that persists in the gas as it cools and recombines (Johnson et al. 2007; Nagakura \& Omukai 2005; Yoshida et al. 2007). The efficient formation of HD can also take place when the primordial gas is collisionally ionized, such as behind the shocks driven by the first SNe or in the virialization of massive DM halos (Greif \& Bromm 2006; Johnson \& Bromm 2006; Machida et al. 2005; Shchekinov \& Vasiliev 2006). In Figure~1, we show the HD fraction in primordial gas in four distinct situations: within a minihalo in which the gas is never strongly ionized, behind a $100~\rm{km}~\rm{s}^{-1}$ shock wave driven by a SN, in the virialization of a $3\sigma$ DM halo at redshift $z=15$, and in the relic H~{\sc ii} region generated by a Pop~III.1 star at $z\sim 20$ (Johnson \& Bromm 2006). Also shown is the critical HD fraction necessary to allow the primordial gas to cool to the temperature floor set by the CMB at these redshifts. Except for the situation of the gas in the virtually un-ionized minihalo, the fraction of HD becomes large quickly enough to play an important role in the cooling of the gas, allowing the formation of Pop~III.2 stars.

\begin{figure}
\centering
\includegraphics[width=.5\textheight]{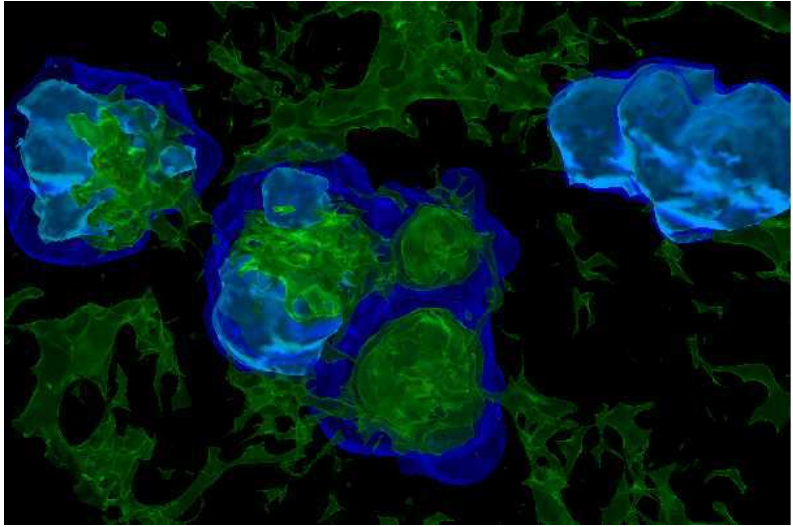}
\caption{The chemical interplay in relic H~{\sc ii} regions. While all molecules are destroyed in and around active H~{\sc ii} regions, the high residual electron fraction in relic H~{\sc ii} regions catalyzes the formation of an abundance of H$_2$ and HD molecules. The light and dark shades of blue denote regions with a free electron fraction of $5\times 10^{-3}$ and $5\times 10^{-4}$, respectively, while the shades of green denote regions with an H$_2$ fraction of $10^{-4}$, $10^{-5}$, and $3\times 10^{-6}$, in order of decreasing brightness. The regions with the highest molecule abundances lie within relic H~{\sc ii} regions, which thus play an important role in subsequent star formation, allowing molecules to become shielded from photodissociating radiation and altering the cooling properties of the primordial gas.}
\end{figure}

Figure~2 schematically shows the characteristic masses of the various stellar populations that form in the early universe. In the wake of Pop~III.1 stars formed in DM minihalos, Pop~III.2 star formation ensues in regions which have been previously ionized, typically associated with relic H~{\sc ii} regions left over from massive Pop~III.1 stars collapsing to black holes, while even later, when the primordial gas is locally enriched with metals, Pop~II stars begin to form (Bromm \& Loeb 2003; Greif \& Bromm 2006). Recent simulations confirm this picture, as Pop~III.2 star formation ensues in relic H~{\sc ii} regions in well under a Hubble time, while the formation of Pop~II stars after the first SN explosions is delayed by more than a Hubble time (Greif et al. 2007; Yoshida et al. 2007a,b; but see Whalen et al. 2008).

\section{Radiative Feedback from the First Stars}
Due to their extreme mass scale, Pop~III.1 stars emit copious amounts of ionizing radiation, as well as a strong flux of H$_2$-dissociating Lyman-Werner (LW) radiation (Bromm et al. 2001b; Schaerer 2002). Thus, the radiation from the first stars dramatically influences their surroundings, heating and ionizing the gas within a few kpc (physical) around the progenitor, and destroying the H$_2$ and HD molecules locally within somewhat larger regions (Alvarez et al. 2006; Abel et al. 2007; Ferrara 1998; Johnson et al. 2007; Kitayama et al. 2004; Whalen et al. 2004). Additionally, the LW radiation emitted by the first stars could propagate across cosmological distances, allowing the build-up of a pervasive LW background radiation field (Haiman et al. 2000).

\subsection{Local Radiative Effects}

The impact of radiation from the first stars on their local surroundings has important implications for the numbers and types of Pop~III stars that form. The photoheating of gas in the minihalos hosting Pop~III.1 stars drives strong outflows, lowering the density of the primordial gas and delaying subsequent star formation by up to $100~\rm{Myr}$ (Johnson et al. 2007; Whalen et al. 2004; Yoshida et al. 2007a). Furthermore, neighboring minihalos may be photoevaporated, delaying star formation in such systems as well (Ahn \& Shapiro 2007; Greif et al. 2007; Shapiro et al. 2004; Susa \& Umemura 2006; Whalen et al. 2007). The photodissociation of molecules by LW photons emitted from local star-forming regions will, in general, act to delay star formation by destroying the main coolants that allow the gas to collapse and form stars.

\begin{figure}
\centering
\includegraphics[width=.35\textheight]{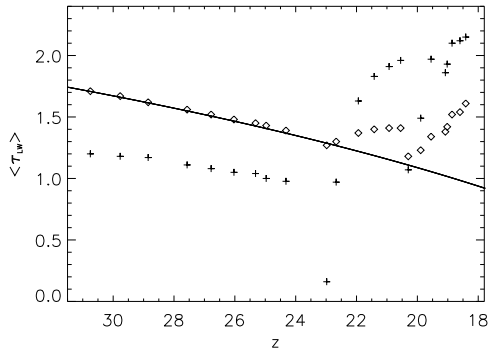}
\caption{Optical depth to LW photons due to self-shielding, averaged over two different scales, as a function of redshift. The diamonds denote the optical depth averaged over the entire cosmological box of comoving length $660~\rm{kpc}$, while the plus signs denote the optical depth averaged over a cube of $220~\rm{kpc}$ (comoving) per side, centered on the middle of the box. The solid line denotes the average optical depth that would be expected for a constant H$_2$ fraction of $2\times 10^{-6}$ (primordial gas), which changes only due to cosmic expansion.}
\end{figure}

The photoionization of primordial gas, however, can ultimately lead to the production of copious amounts of molecules within the relic H~{\sc ii} regions surrounding the remnants of Pop~III.1 stars (Johnson \& Bromm 2007; Nagakura \& Omukai 2005; Oh \& Haiman 2002; Ricotti et al. 2001). Recent simulations tracking the formation of, and radiative feedback from, individual Pop~III.1 stars in the early stages of the assembly of the first galaxies have demonstrated that the accumulation of relic H~{\sc ii} regions has two important effects. First, the HD abundance that develops in relic H~{\sc ii} regions allows the primordial gas to re-collapse and cool to the temperature of the CMB, possibly leading to the formation of Pop~III.2 stars in these regions (Johnson et al. 2007; Yoshida et al. 2007b). Second, the molecule abundance in relic H~{\sc ii} regions, along with their increasing volume-filling fraction, leads to a large optical depth to LW photons over physical distances of the order of several kpc. The development of a high optical depth to LW photons over such short length-scales suggests that the optical depth to LW photons over cosmological scales may be very high, acting to suppress the build-up of a background LW radiation field, and mitigating negative feedback on star formation.

Figure~3 shows the chemical composition of primordial gas in relic H~{\sc ii} regions, in which the formation of H$_2$ molecules is catalyzed by the high residual electron fraction. Figure~4 shows the average optical depth to LW photons across the simulation box, which rises with time owing to the increasing number of relic H~{\sc ii} regions.

\begin{figure}
\centering
\includegraphics[width=.65\textheight]{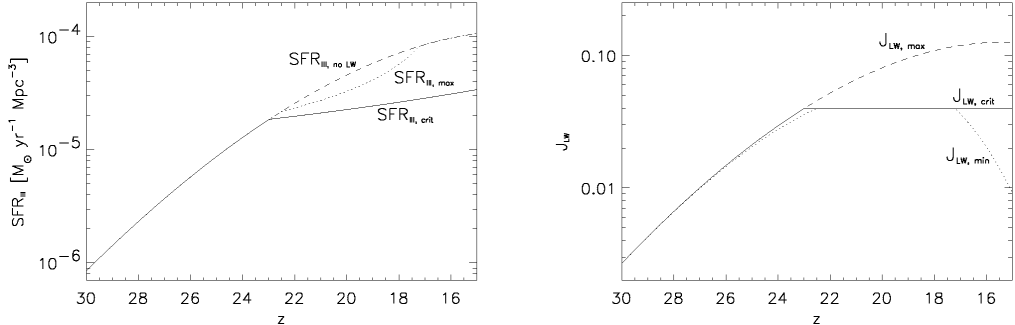}
\caption{The Pop III star formation rates ({\it left panel}) and the corresponding LW background fluxes ({\it right panel}) for three models of the build-up of the LW background by Pop III.1 stars formed in minihalos.  
The maximum possible LW background, $J_{\rm LW, max}$, is generated for the case that every minihalo with a virial temperature $\geq$ 2 $\times$ 10$^3$ K hosts a Pop III star, without the LW background in turn diminishing the SFR, labeled here as SFR$_{\rm III, no LW}$. The self-regulated model considers the coupling between the star formation rate, SFR$_{\rm LW, crit}$, and the critical LW background that it produces, $J_{\rm LW, crit}$. The minimum value for the LW background, $J_{\rm LW, min}$, is produced for the case of a high opacity through the relic H~II regions left by the first stars, in which case the self-consistent SFR, SFR$_{\rm III, max}$ can approach the undiminished SFR$_{\rm III, no LW}$. }
\end{figure}

\subsection{Global Radiative Feedback}
While the reionization of the universe is likely to have occurred at later times, as inferred from the {\it WMAP} third year results (Spergel et al. 2007), the process of primordial star formation can be affected by the build up of a LW background very soon after the formation of the first stars.  This LW radiation, which acts to destroy H$_2$, the very coolant that enables the formation of the first stars, can, in principle, dramatically lower the formation rate of Pop III stars in minihalos (e.g. Haiman et al. 2000; Machacek et al. 2001; Yoshida et al. 2003; Mackey et al. 2003).  

While star formation in more massive systems may proceed relatively unimpeded, through atomic line cooling, during the earliest epochs of star formation these atomic-cooling halos are rare compared to the minihalos which host individual Pop III stars.  While the process of star formation in atomic-cooling halos is not well understood, for a broad range of models the dominant contribution to the LW background is from Pop III.1 stars formed in minihalos at $z$ $\geq$ 15-20 (Johnson et al. 2008).  Therefore, at these redshifts the LW background radiation may be largely self-regulated, with Pop III.1 stars producing the very radiation which, in turn, suppresses their formation.  Johnson et al. (2008) argue that there is a critical value for the LW flux, $J_{\rm LW, crit}$ $\sim$ 0.04, at which Pop III.1 star formation occurs self-consistently, with the implication that the PopIII.1 star formation rate in minihalos at $z$ $>$ 15 is decreased by only a factor of a few, as shown in Fig. 5.  The star formation rate may be even higher if the cosmological average optical depth to LW photons through the relic H~II regions left by the first stars is sufficiently high.  An analytical model of this effect shows that the SFR may be only negligibly reduced once the volume-filling factor of relic H~II regions becomes large, as is also shown in Fig. 5 (Johnson et al. 2008).  

Simulations of the formation of a dwarf galaxy at $z$ $\geq$ 10 which take into account the effect of a LW background at $J_{\rm LW, crit}$ show that Pop III.1 star formation takes place before the galaxy is fully assembled, suggesting that the formation of metal-free galaxies may be a rare event in the early universe (Johnson et al. 2008).  Figure 6 shows the temperature and density of the protogalaxy simulated by these authors at $z \sim 12.5$.     

\begin{figure}
\centering
\includegraphics[width=.55\textheight]{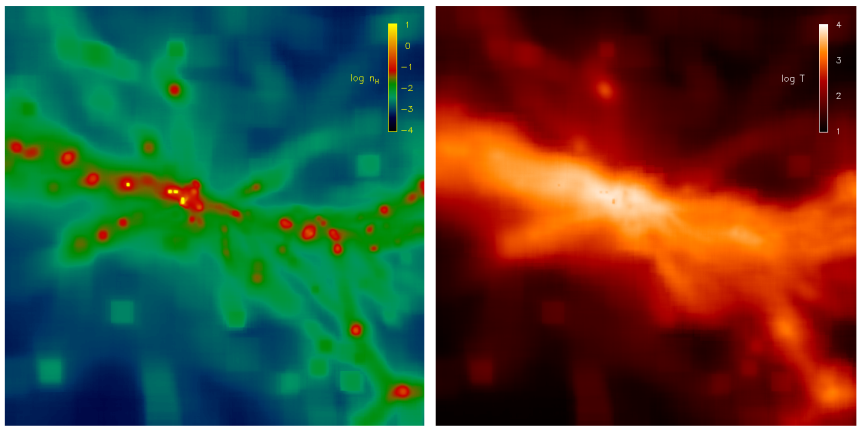}
\caption{The hydrogen number density and temperature of the gas in the region of the forming galaxy at $z$ $\sim$ 12.5.  
The panels show the inner $\sim $ 10.6 kpc (physical) of our cosmological box.   The cluster of minihaloes harboring dense 
gas just left of the center in each panel is the site of the formation of the two Pop III.1 stars which are able to form 
in our simulation including the effects of the self-regulated LW background.  The remaining minihaloes are not able to form 
stars by this redshift, largely due to the photodissociation of the coolant H$_2$.  The main progenitor DM halo, which hosts 
the first star at $z$ $\sim$ 16, by $z$ $\sim$ 12.5 has accumulated a mass of 9 $\times$ 10$^7$ M$_{\odot}$ through 
mergers and accretion.  Note that the gas in this halo has been heated to temperatures above 10$^4$ K,  leading to a high 
free electron fraction and high molecule fractions in the collapsing gas. The high HD fraction that is 
generated likely leads to the formation of Pop III.2 stars in this system. }
\end{figure}

\section{The First Supernova Explosions}
Recent numerical simulations have indicated that primordial stars forming in DM minihalos typically attain $100~M_{\odot}$ by efficient accretion, and might even become as massive as $500~M_{\odot}$ (Bromm \& Loeb 2004; O'Shea \& Norman 2007; Omukai \& Palla 2003; Yoshida et al. 2006). After their main-sequence lifetimes of typically $2-3~\rm{Myr}$, stars with masses below $\simeq 100~M_{\odot}$ are thought to collapse directly to black holes without significant metal ejection, while in the range $140-260~M_{\odot}$ a pair-instability supernova (PISN) disrupts the entire progenitor, with explosion energies ranging from $10^{51}-10^{53}~\rm{ergs}$, and yields up to $0.5$ (Heger et al. 2003; Heger \& Woosley 2002). 

\begin{figure}
\centering
\includegraphics[width=.45\textheight]{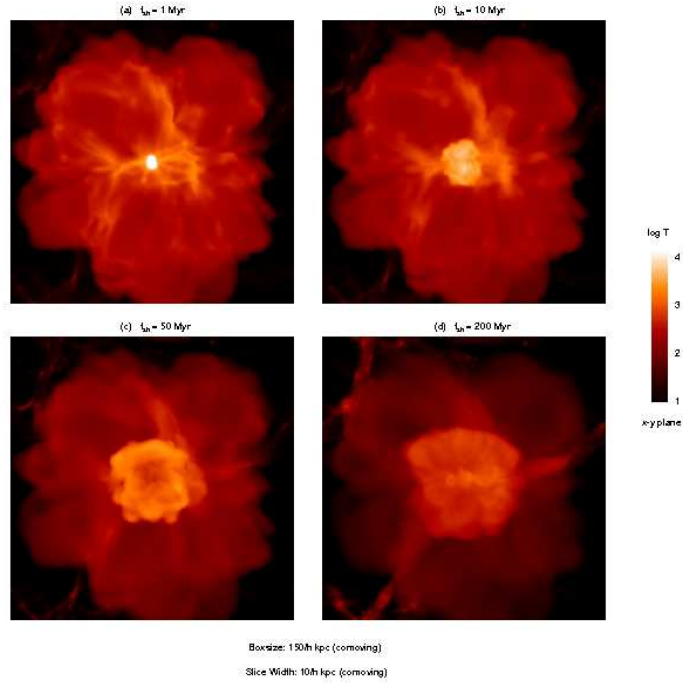}
\caption{Temperature averaged along the line of sight in a slice of $10/h~\rm{kpc}$ (comoving) at $1$,  $10$, $50$, and $200~\rm{Myr}$ after a Pop III PISN (Greif et al. 2007). In all four panels, the H~{\sc ii} region and SN shock are clearly distinguishable, with the former occupying almost the entire simulation box, while the latter is confined to the central regions. {\it (a)}: The SN remnant has just left the host halo, but temperatures in the interior are still well above $10^{4}~\rm{K}$. {\it (b)}: After $10~\rm{Myr}$, the asymmetry of the SN shock becomes visible, while most of the interior has cooled to well below $10^{4}~\rm{K}$. {\it (c)}: The further evolution of the shocked gas is governed by adiabatic expansion. {\it (d)}: After $200~\rm{Myr}$, the shock velocity approaches the local sound speed and the SN remnant stalls. By this time the post-shock regions have cooled to roughly $10^{3}~\rm{K}$.}
\end{figure}

The significant mechanical and chemical feedback effects exerted by such explosions have been investigated with a number of detailed calculations, but these were either performed in one dimension (Kitayama \& Yoshida 2005; Machida et al. 2005; Salvaterra et al. 2004; Whalen et al. 2008), or did not start from realistic initial conditions (Bromm et al. 2003; Norman et al. 2004). The most realistic, three-dimensional simulation to date took cosmological initial conditions into account, and followed the evolution of the gas until the formation of the first minihalo at $z\simeq 20$ (Greif et al. 2007). After the gas approached the `loitering regime' at $n_{\rm{H}}\simeq 10^{4}~\rm{cm}^{-3}$, the formation of a primordial star was assumed, and a photoheating and ray-tracing algorithm determined the size and structure of the resulting H~{\sc ii} region (Johnson et al. 2007). An explosion energy of $10^{52}~\rm{ergs}$ was then injected as thermal energy into a small region around the progenitor, and the subsequent expansion of the SN remnant was followed until the blast wave effectively dissolved into the IGM. The cooling mechanisms responsible for radiating away the energy of the SN remnant, the temporal behavior of the shock, and its morphology could thus be investigated in great detail (see Figure~7).

\begin{figure}
\centering
\includegraphics[width=.35\textheight]{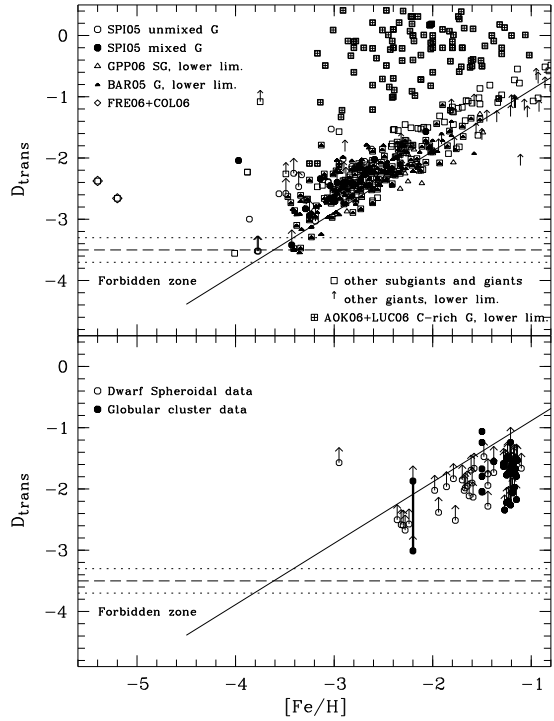}
\caption{Transition discriminant, $D_{\rm{trans}}$, for metal-poor stars collected from the literature as a function of [Fe/H]. {\it Top panel}: Galactic halo stars. {\it Bottom panel}: Stars in dSph galaxies and globular clusters. G indicates giants, SG subgiants. The critical limit is marked with a dashed line. The dotted lines refer to the uncertainty. The detailed references for the various data sets can be found in Frebel et al. (2007).}
\end{figure}

The dispersal of metals by the first SN explosions transformed the IGM from a simple primordial gas to a highly complex medium in terms of chemistry and cooling, which ultimately enabled the formation of the first low-mass stars. However, this transition required at least a Hubble time, since the presence of metals became important only after the SN remnant had stalled and the enriched gas re-collapsed to high densities (Greif et al. 2007; but see Whalen et al. 2008). Furthermore, the metal distribution was highly anisotropic, as the post-shock gas expanded into the voids in the shape of an `hour-glass', with a maximum extent similar to the final mass-weighted mean shock radius (Greif et al. 2007).

To efficiently mix the metals with all components of the swept-up gas, a DM halo of at least $M_{\rm{vir}}\simeq 10^{8}~M_{\odot}$ had to be assembled (Greif et al. 2007), and with an initial yield of $0.1$, the average metallicity of such a system would accumulate to $Z\simeq 10^{-2.5}Z_{\odot}$, well above any critical metallicity (Bromm \& Loeb 2003; Bromm et al. 2001a; Schneider et al. 2006; see also Wise \& Abel 2007). Thus, if energetic SNe were a common fate for the first stars, they would have deposited metals on large scales before massive galaxies formed and outflows were suppressed. Hints to such ubiquitous metal enrichment have been found in the low column density Ly$\alpha$ forest (Aguirre et al. 2005; Songaila \& Cowie 1996; Songaila 2001), and in dwarf spheroidal satellites of the Milky Way (Helmi et al. 2006).

\section{The Chemical Signature of the First Stars}

The discovery of extremely metal-poor stars in the Galactic halo has made studies of the chemical composition of low-mass Pop~II stars powerful probes of the conditions in which the first low-mass stars formed. While it is widely accepted that metals are required for the formation of low-mass stars, two general classes of competing models for the Pop~III -- Pop~II transition are discussed in the literature: {\it (i)} atomic fine-structure line cooling (Bromm \& Loeb 2003; Santoro \& Shull 2006); and {\it (ii)} dust-induced fragmentation (Schneider et al. 2006). Within the fine-structure model, C~{\sc ii} and O~{\sc i} have been suggested as main coolants (Bromm \& Loeb 2003), such that low-mass star formation can occur in gas that is enriched beyond critical abundances of $\mbox{[C/H]}_{\rm{crit}}\simeq -3.5\pm 0.1$ and $\mbox{[O/H]}_{\rm{crit}}\simeq -3\pm 0.2$. The dust-cooling model, on the other hand, predicts critical abundances that are typically smaller by a factor of $10-100$.

Based on the theory of atomic line cooling (Bromm \& Loeb 2003), a new function, the `transition discriminant' has been introduced:
 \begin{equation}
 D_{\rm{trans}}\equiv \log_{10}\left(10^{\mbox{[C/H]}}+0.3\times 10^{\mbox{[O/H]}}\right)\mbox{\ ,}
 \end{equation}
such that low-mass star formation requires $D_{\rm{trans}}>D_{\rm{trans,crit}}\simeq -3.5\pm 0.2$ (Frebel et al. 2007). Figure~8 shows values of $D_{\rm{trans}}$ for a large number of the most metal-poor stars available in the literature. While theories based on dust cooling can be pushed to accommodate the lack of stars with $D_{\rm{trans}}<D_{\rm{trans,crit}}$, it appears that the atomic-cooling theory for the Pop III -- Pop II transition naturally explains the existing data on metal-poor stars. Future surveys of Galactic halo stars will allow to further populate plots such as Figure~8, and will provide valuable insight into the conditions of the early universe in which the first low-mass stars formed.

The abundance patterns observed in the most metal-poor stars can also provide information about the types of SN that ended the lives of the first stars, as the metals that are emitted in these explosions will become incorporated into later generations of stars, some of which are observed in the halo of the Galaxy.  Interestingly, while detailed numerical simulations of the formation of the first stars suggest that they were often massive enough to explode as PISN, no clear signature of such a PISN has yet been detected in a metal-poor star.  Does this imply that the first stars did not explode as PISN, or that they were not very massive (see also Ekstrom in this proceedings)?  

\begin{figure}
\centering
\includegraphics[width=.4\textheight]{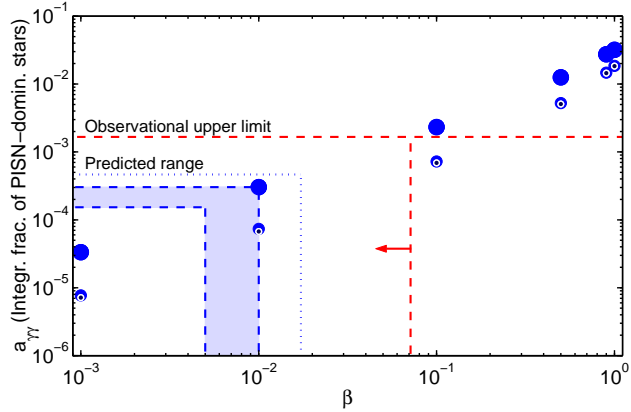}
\caption{The predicted integrated (total) fraction of PISN-dominated stars below $[\mathrm{Ca}/\mathrm{H}]=-2$ as a function of $\beta$, corresponding to $>90\%$ (big blue dots), $>99\%$ (medium sized, blue dots), and $>99.9\%$ (small, dark blue dots) PISN-enrichment.  The dashed (red) lines indicate the observational upper limit of $\beta$, assuming that none of the $\sim 600$ Galactic halo stars with $[\mathrm{Ca}/\mathrm{H}]\le -2$ for which high-resolution spectroscopy is available show any signature of PISNe (N. Christlieb, priv. comm.).  The dotted (blue) line and shaded (blue) area denote the predicted range of $a_{\gamma\!\gamma}$ anticipated from the calculations of Padoan et al. (2007) and Greif \& Bromm (2006), respectively.  }
\end{figure}

PISNe may have ejected enough mass in metals to enrich the IGM to a metallicity well above those of the most metal-poor stars. 
Therefore, one possible explanation for the apparent lack of Pop III PISNe is that the few stars which might have formed from gas dominantly enriched by a PISN may have relatively high metallicities, and so may have eluded surveys seeking such true second generation stars at lower metallicity (Karlsson et al. 2008).  Karlsson et al. (2008) developed a model for the inhomogeneous chemical enrichment of the gas collapsing to become a dwarf galaxy at z $\sim$ 10 in which the formation of both Pop III stars from metal-free gas and the formation of Pop II stars from metal-enriched gas were tracked self-consistently.
These authors find that the lack of the discovery of a metal-poor star showing signs of enrichment dominated by PISN yields in the existing catalog of metal poor stars is not inconsistent with theories predicting that the first stars were very massive, as shown in Figure 9. It is hoped that future surveys of stars in the Galactic halo will test this model by searching for PISN-enriched stars with metallicities [Fe/H] $\geq$ -2.5.

\section{Conclusion}
Understanding the formation of the first galaxies marks the frontier of high-redshift structure formation. It is crucial to predict their properties in order to develop the optimal search and survey strategies for the {\it JWST}. Whereas {\it ab-initio} simulations of the very first stars can be carried out from first principles, and with virtually no free parameters, one faces a much more daunting challenge with the first galaxies. Now, the previous history of star formation has to be considered, leading to enhanced complexity in the assembly of the first galaxies. One by one, all the complex astrophysical processes that play a role in more recent galaxy formation appear back on the scene. Among them are external radiation fields, comprising UV and X-ray photons, and possibly cosmic rays produced in the wake of the first SNe (Stacy \& Bromm 2007). There will be metal-enriched pockets of gas which could be pervaded by dynamically non-negligible magnetic fields, together with turbulent velocity fields built up during the virialization process. However, the goal of making useful predictions for the first galaxies is now clearly drawing within reach, and the pace of progress is likely to be rapid.

\begin{acknowledgments}

\end{acknowledgments}
V.B. acknowledges support from NSF grant AST-0708795. The simulations 
presented here were carried out at the Texas Advanced Computing Center (TACC).

\end{document}